\documentclass[twocolumn,amsmath,amssymb,aps,physrev]{revtex4-2}
 
\usepackage{ulem}
\usepackage{amsmath}
\usepackage{multirow}
\usepackage{graphicx}
\usepackage{dcolumn}
\usepackage{bm}

\begin{document}

\preprint{APS/123-QED}

\title{Unified Description of Kappa-type velocity distributions}
\author{J. A. S. Lima$^{1}$} \email{jas.lima@iag.usp.br}  
\author{M. H. Benetti$^{1}$} \email{mhbenetti@usp.br}

\affiliation{$^{1}$Departamento de Astronomia (IAG-USP), Universidade de S\~ao Paulo, 05508-090 S\~ao Paulo SP,  Brasil}

\date{\today}

\begin{abstract}
 An extension of Maxwell's original prescription for an ideal gas is adopted to derive a broad class of Kappa-type velocity distributions, encompassing both fat and short-tailed forms. Within this general framework, a physically consistent fat-tailed Kappa distribution is identified that accurately fits recent suprathermal data. In particular, a kinetic physical temperature $T$ emerges naturally from the model, eliminating the need to invoke an effective temperature $T_{\kappa\ell}$, as is commonly done in the literature. Finally, it is argued that only a particular value of $\ell$ ensures a satisfactory fit to the data when the physical kinetic temperature is employed.

\end{abstract}


\maketitle


\noindent{\it 1. Introduction.} Several types of velocity distribution functions (VDFs), collectively dubbed $\kappa$-distributions, are employed in numerous applications involving space plasmas across diverse environments
\cite{Marsch2006,Christon1987,Ogasawara2017}. A large set of independent observations have shown that suprathermal particles in their tails cannot be described by a Maxwellian distribution. Of particular interest are the data collected by Voyager spacecraft \cite{Decker2003}, HELIOS  \cite{Pierrard2010}, WIND \cite{Salem2023}, Cluster satellites \cite{Silveira2021}, STEREO \cite{Wang2012}, Juno mission \cite{McComas2017}, and Parker Solar Probe \cite{Benetti2023}. From a theoretical perspective, after the first phenomenological suggestion of a kappa-type distribution \cite{Vasyliunas1968,Olbert1968}, several authors proposed slightly altered kappa versions \cite{Hasegawa1985,Maksimovic1997,Scherer2018}.

Table \ref{T0} displays for $N$ particles in a volume $V$, four different 3D Kappa-type fat-tailed distributions (not in historical order), where $n=N/V$ is the concentration and $\Gamma$ is the Gamma function. The quantities $v_{te}$, $\omega_0$, $\theta$, $v_{\kappa}$,  represent typical velocities, whose definitions depend on the chosen formulation. Such quantities are, respectively, part of the first kind of Kappa distribution (FKK)\cite{Hasegawa1985}, second kind of Kappa (SKK)\cite{Vasyliunas1968}, third kind of Kappa (TKK)\cite{Livadiotis2013}, and the modified second kind of Kappa distribution (MSKK)\cite{Maksimovic1997} commonly referred to as the modified Kappa distribution (MK) in the literature\cite{Lazar2022}. 


The FKK and SKK types were named by Livadiotis \& McComas\cite{Livadiotis2009}. The TKK emerges in the context of the Hamiltonian distribution function, while MSKK is a modification of the SKK distribution by assuming that the temperature of the suprathermal gas depends on $\kappa$. 

In this article, a general $\kappa$-distribution is derived through an extended\, ``Neo-Maxwellian" approach, inspired by Maxwell's seminal work\cite{Maxwell1860}. All distributions in Table \ref{T0} will be described by a pair of free parameters ($b,\ell$). The first is related to the $\kappa$-index and governs the deformation relative to its Gaussian limit, while $\ell$ is a pure number in the exponent (for instance, 0, -1 and -5/2 in Table I). As we shall see, apart from the $\ell=0$ case, such values of $\ell$ constrain the power law to have the same expression of a given Maxwellian quantity. In principle, such results allow us to pick the most realistic distribution from a physical point of view. We identify a class of power-law distributions in which the mean energy, and hence the temperature, is independent of the deformation parameter. This remarkable property eliminates the need for effective temperatures and may help resolve persistent ambiguities in defining temperature for fat-tailed systems.


\noindent\textit{2. Deformation of Euler Exponential and VDFs.}
To begin with, let us consider a two-parametric deformation of the Euler exponential relation:
\begin{equation}\label{E2}
e_{b\ell}(f) =  \left( 1 + bf \right)^{\frac{1}{b} + \ell}\equiv \left( 1 + bf \right)^{\frac{1+b\ell}{b}}, 
\end{equation}
where  $b$ is a convenient deformation parameter, while $\ell$ is a real finite number. 
For $\ell \neq -1/b$, one can see that $\lim_{b\rightarrow 0}\, e_{b\ell}(f) = \exp(f)$. We also introduce the deformed $(b\,\ell)$-logarithm, which is the inverse function of $e_{b\ell}(f)$:
\begin{equation}\label{E3}
    \ln_{b\ell}(f) = \frac{f^{\frac{b}{1 + b\ell}} - 1}{b}.
\end{equation}
As one may check, $\ln_{b\ell}(e_{b\ell}f)= e_{b\ell}(\ln_{b\ell}f)\equiv f$. 
\begin{table}[t!]
\renewcommand{\arraystretch}{2.8}
\caption{\label{T0} Fat-tailed Kappa-type distributions: FKK $(i)$, SKK $(ii)$, TKK $(iii)$ and MSKK $(iv)$.}
\begin{ruledtabular}
\begin{tabular}{l}
\makebox[\linewidth][l]{%
$(i)\quad \quad F(v) = \displaystyle n\left(\frac{1}{2\pi v_{te}^2 }\right)^{3/2} \frac{ \Gamma\left(\kappa  \right)}{\kappa^{3/2} \Gamma\left({\kappa}- \frac{3}{2} \right)}\left[1 +  \frac{ v^2}{2\kappa v_{te}^2}\right]^{-\kappa}$} \\[3pt]
\makebox[\linewidth][l]{%
$(ii)\quad\,\,\,\, F(v) =\displaystyle n\left(\frac{1}{\pi \omega_0^2}\right)^{3/2} \frac{ \Gamma\left(\kappa  +1\right)}{\kappa^{3/2} \Gamma\left({\kappa} -\frac{1}{2} \right)}\left[1 +  \frac{ v^2}{\kappa \omega_0^2}\right]^{-\kappa-1}$} \\[3pt]
\makebox[\linewidth][l]{%
$(iii)\quad \,\,F(v) = \displaystyle n\left(\frac{1}{\pi \theta^2}\right)^{3/2} \frac{ \Gamma\left(\kappa + \frac{5}{2}\right)}{\kappa^{3/2} \Gamma\left({\kappa} +1 \right)}\left[1 +  \frac{ v^2}{\kappa \theta^2}\right]^{-\kappa-\frac{5}{2}}$}\\ [5pt]
$(iv)\quad\,\,\, F(v) =\displaystyle n \displaystyle\left( \frac{1}{\pi  v_{\kappa}^2} \right)^{3/2} \frac{ \Gamma\left(\kappa+ 1\right)}{ \kappa^{3/2}\Gamma\left({\kappa}-\frac{1}{2}\right)} \left[1 +  \frac{ v^2}{ \kappa v_{\kappa}^2}\right]^{-\kappa - 1} $ \\[5pt] 
\end{tabular}
\end{ruledtabular}
\end{table}
We briefly note that the exponential introduced in Eq. (\ref{E2}) defines a generalized class of velocity distributions derived by following Maxwell’s well-established prescription for determining equilibrium. This approach allows for systematic extensions beyond the Gaussian case, including several relevant fat-tailed distributions. In Maxwell's formulation\cite{Maxwell1860}, it was hypothesized that the velocity components are statistically independent, leading to $F(v)=f(v_x)f(v_y)f(v_z)$, a rule of thumb for Gaussian distributions. Here we assume  
\begin{equation}\label{E4}
F(v) = e_{b\ell} \left\{ \ln_{b\ell}[f(v_x)] + \ln_{b\ell}[f(v_y)]+ \ln_{b\ell}[f(v_z)]\right\}. 
\end{equation}
Note that Maxwell's kinetic hypothesis is always recovered in the limit $b\rightarrow 0$. 
Moreover, we also adopt the assumption of isotropy in velocity space $F(v)= F(\sqrt{v_x^{2}+ v_x^{2} + v_x^{2}})$. So, by taking the deformed logarithm on both sides of (\ref{E4}) and differentiating with respect to $v_x$, $v_y$, and $v_z$ yields the following
\begin{eqnarray}\label{E5}
\Phi(v) &=& \frac{1}{v_x}\frac{d \left[\ln_{b\ell}f(v_x)\right] }{d v_x} = \frac{1}{v_y}\frac{d\left[ \ln_{b\ell}f(v_y) \right]}{d v_y} \nonumber \\
&=& \frac{1}{v_z}\frac{d \left[\ln_{b\ell}f(v_z) \right]}{d v_z},
\end{eqnarray}
where we have defined: 
\begin{equation}\label{E6}
\Phi(v) = \frac{F'(v)}{v}\frac{d \left[\ln_{b\ell} F(v)\right]}{d F(v)}.
\end{equation}
The equalities in (\ref{E5}) means that each term must be constant but may depend both on the mass $m$ and the temperature $T$. In the one-dimensional case, one finds:
\begin{equation}\label{E7}
\frac{1}{v_x}\frac{d \ln_{b\ell}f(v_x) }{d v_x} = -\beta_{b\ell}\, m, 
\end{equation}
where $\beta_{b\ell}$ incorporates the deformation parameters and the thermodynamic properties of the gas system. Integrating Eq. (\ref{E7}) we have
\begin{equation}\label{E8} 
\ln_{b\ell}f(v_x)  = -\frac{\beta_{b\ell}\, m v_x^2}{2} + \ln_{b\ell}A_1 , 
\end{equation}
with $A_1$ being a normalization constant for the 1-D distribution function. Extending this to three dimensions, we sum the contributions from all components, thus, the deformed logarithm of Eq. (\ref{E4}) can be written as
\begin{equation}\label{E9}
\ln_{b\ell} F(v)  = -\frac{\beta_{b\ell} \,m \left(v_x^2+v_y^2+v_z^2\right)}{2} + \ln_{b\ell} A_3, 
\end{equation}
where  $A_3$ is new constant. Now, using the identity:
\begin{equation}\label{E10}
\ln_{b\ell}F(v)- \ln_{b\ell}A_3 = A_3^{\frac{d}{1+d\ell}}\,\,\ln_{b\ell}\left[\frac{F(v)}{A_3}\right],
\end{equation}
in Eq. (\ref{E9}), we see that
\begin{equation}\label{E12}
\ln_{b\ell}\left[\frac{F(v)}{A_3}\right] = -{\beta_{b\ell}}{A_3^{-\frac{d}{1+d\ell}}}\frac{m \left(v_x^2+v_y^2+v_z^2\right)}{2}. 
\end{equation}

Introducing the inverse of the typical energy scale, 
\begin{equation}\label{E13}
\beta_* = {\beta_{b\ell}}{A_3^{-\frac{d}{1+d\ell}}}, 
\end{equation}
we compute the deformed exponential in Eq. (\ref{E12})  and normalize it to the concentration $n$.  Hence, for $b<0 \,\,(b \to -|b|)$ and $b>0$, we obtain the 3-D generalized fat and short-tailed distributions, respectively:

\begin{subequations}\label{E14}
\begin{align}
   F(v) &= n\left(\frac{\beta_* m}{2\pi }\right)^{\frac{3}{2}} \frac{|b|^{\frac{3}{2}} \Gamma\left(\frac{1}{|b|} - \ell\right)}{ \Gamma\left(\frac{1}{|b|} - \ell - \frac{3}{2}\right)}\left[1 + \frac{|b|\beta_* m v^2}{2}\right]^{-\frac{1}{|b|} + \ell}  \label{E14a} \\[10pt]
   F(v) &= n \left(\frac{\beta_* m }{2\pi }\right)^{\frac{3}{2}} \frac{b^{\frac{3}{2}} \Gamma\left(\frac{1}{b}+\ell+ \frac{5}{2}\right)}{\Gamma\left( \frac{1}{b}+\ell+ 1 \right)}\left[1 - \frac{b \beta_* m v^2}{2}\right]^{\frac{1}{b} + \ell}. \label{E14b}
\end{align}
\end{subequations} 
\begin{table}[t!]
\renewcommand{\arraystretch}{2.7}
\caption{\label{T1} Physical quantities for Fat-tailed distributions}
\begin{ruledtabular}
\begin{tabular}{lc}
\noalign{\vskip -5pt}
\,\,\,\,\,\,\, \textrm{Fat-Tail ($b < 0$)} & \,\textrm{Maxwell (M) limits ($b\to0$)} \\
\colrule
$\bar{\varepsilon}_{b\ell}=\displaystyle\bar{\varepsilon}_M \left( \frac{1}{1- 5|b|/2-|b|\ell } \right)$ & $\displaystyle\bar{\varepsilon}_M = \displaystyle\frac{3k_B T}{2}$ \\[5pt]
$P_{b\ell}=\displaystyle P_M \left(\frac{1}{1- 5|b|/2-|b|\ell }\right)$ & $\displaystyle P_M = \displaystyle nk_B T$ \\ [5pt]
$v^{\text{rms}}_{b\ell}=\displaystyle v^{\text{rms}}_M\sqrt{\frac{1}{1- 5|b|/2-|b|\ell }} $ & $ \displaystyle\displaystyle v^{\text{rms}}_M=\sqrt{\frac{3k_B T}{m}} $ \\[5pt]
$\bar{v}_{b\ell}= \displaystyle\bar{v}_M \frac{\Gamma\left(\frac{1}{|b|}-\ell -2\right)}{|b|^{1/2} \Gamma\left(\frac{1}{|b|}-\ell -\frac{3}{2}\right)} $ & $ \displaystyle \bar{v}_M=\sqrt{\frac{8k_B T}{\pi m}} $ \\[6pt]
$v^{\text{mp}}_{b\ell}= \displaystyle v^{\text{mp}}_M\sqrt{\frac{1}{1-|b|-|b|\ell}} $ & $ \displaystyle v^{\text{mp}}_M=\sqrt{\frac{2k_B T}{m}} $ \\[5pt]
\end{tabular}
\end{ruledtabular}
\end{table}
As should be expected, these $F(v)$ expressions reduce to the same Gaussian classical limit \cite{NoteEulerLimit}:
\begin{equation}\label{E16}
    \lim_{b \to 0}F(v) = n\left(\frac{\beta_* m}{2\pi}\right)^{3/2} \exp\left(-\frac{\beta_* m v^2}{2}\right).
\end{equation}
Hence, the Maxwellian distribution is recovered only if $\beta_*=\beta=(k_BT)^{-1}$, where $k_B$ is the Boltzmann constant. As a result, the method adopted here allowed us to isolate the possible $\beta$-dependence on the pair ($b,\ell$) for all power laws defined by (\ref{E14a}) and (\ref{E14b}). Henceforth, $\beta$  will assume its standard value, and as such, the zeroth law of thermodynamics (thermal equilibrium) is naturally obeyed for the entire class of power laws deduced here.

Table \ref{T1} displays the kinetic averaged scales (energy per particle, pressure, and typical velocities) of the fat-tailed distributions modulated by the corresponding Maxwellian values.  When $b \to 0$, all Maxwellian results are recovered. In the nontrivial case ($\bar{v}$), the limiting properties of the $\Gamma$-function provide the same result.

The parameter $\beta = (k_B T)^{-1}$ can also be eliminated from $\bar{\varepsilon}_{b\ell}$ and $P_{b\ell}$ from the fat and short-tailed distributions (see Table II and also Table IV in Appendix A). In addition, by using $n=N/V$ and $\quad U_{b\ell}\equiv N\bar{\varepsilon}_{b\ell}$, a general equation of state (\textbf{EoS}), independent of the pair ($b,\ell)$, is obtained:
\begin{equation}\label{E56}
    \frac{P_{b\ell}V}{U_{b\ell}} =  \frac{P_{M}V}{U_M}=\frac{2}{3},
\end{equation}
where $P_M$ and $U_M$ are the Maxwellian standard expressions. Any ($b,\ell$)-VDF obeys this simple relation. Equation (\ref{E56}) also holds for classical and quantum ideal gases \cite{Landau1985}, as well as for Tsallis' q-statistics \cite{Lima2005}. 

For a given $b$ (or $\kappa$), the pressure depends on the values of $\ell$ (see Tables II and IV). However, for $\ell=-5/2$, the common ideal gas law remains valid, that is, $P_{b({\ell}=-5/2)} = nk_BT$, and the same happens with the average energy per particle, $\bar{\varepsilon}_{b ({\ell}=-5/2)} =  3k_BT/2$.\\
\noindent{\it 3. Fat-Tailed Kappa Distributions.} Let us now determine the physical role played by the pair $(b,\ell)$. For a while, we focus our attention on the fat-tailed class.  By comparing Eq. (\ref{E14a}) with Table \ref{T0}, we see that the deformation parameter $b$ of FKK, SKK and TKK can be mapped in the $\kappa$-index by the simple relation $|b| = \kappa^{-1}$. Hence, the two-parametric class of power-law VDF is given by: 
\begin{equation}\label{Eqn24}
     F(v) = n\left(\frac{m}{2\pi k_B T }\right)^{\frac{3}{2}} \frac{ \Gamma\left(\kappa - \ell\right)}{\kappa^{\frac{3}{2}} \Gamma\left(\kappa - \ell - \frac{3}{2}\right)} \left[1 + \frac{ m v^2}{2\kappa k_B T}\right]^{-(\kappa - \ell)}
\end{equation}
This is the ultimate form of the general fat-tailed ($\kappa-\ell$) distribution in the present approach (see Appendix A for short-tailed VDFs).   
From Table I, we see that some values of $\ell$ have already been phenomenologically adopted, namely: $\ell = 0$ (FKK), $\ell = -1$ (SKK) and $\ell = -5/2$ (TKK). The MSKK ($\ell=-1$) VDF in Table I will be discussed separately because it is a special case where an effective temperature $T_{\kappa\,\ell}$, for $\ell =-1$, has been introduced. \\

We stress that the temperature in our approach always appears isolated in the term $2k_B T/m$, which in the Maxwellian limit represents the square of the most probable velocity ($v^{{\text{mp}}}_{M}$) of the gas particles. However, despite the physical temperature $T$ be independent of $\kappa$ and  $\ell$,  when the system follows a power-law distribution, most of the averaged physical quantities, including $v^{{\text{mp}}}_{M}$, may depend on the pair ($\kappa,\,\ell$), and thus requires a case-by-case analysis.

In Table \ref{T1.1}, unless the MSKK case, the final form of the VDFs are shown for the identified values of $\ell = 0,-1,-5/2$.  
For $\ell = 0$ (FKK), we have a $\kappa$-deformation distribution, which has been applied in studies of power law behavior within the non-additive Tsallis q-statistics framework\cite{Tsallis1988}, both in the statistical ensemble and kinetic approaches\cite{Silva1998,Lima2001,Lima2020,Lima2025}. This particular choice unifies the Tsallis and FKK distributions, establishing a connection between the Kappa distribution and the non-extensive formalism as discussed long ago\cite{Leubner2002}. Now, the present two-parametric method shows in a very simple way that the Tsallis VDF is a particular case of (\ref{Eqn24}). In the fat-tailed scenario, all physical quantities depend on $|b| = \kappa^{-1} = q-1$, which implies that none of its typical velocities corresponds to Maxwellian values (cf. Table II). When \( \ell = -1 \),  $v^{\text{mp}}_{\kappa\,\ell}= v^{\text{mp}}_M$. For this reason, the SKK model, originally introduced by Vasyliunas (1968) as the first Kappa-type distribution, correctly assumes $\omega_0$ as the most probable Maxwellian velocity. Nevertheless, for the TKK, $\ell=-5/2$, the most probable velocity is:
\begin{equation}\label{24}
v^{\text{mp}}_{(\text{TKK})} = v^{\text{mp}}_M \displaystyle\sqrt{\frac{\kappa}{\kappa + 3/2}}. 
\end{equation}
Hence, for $\ell=0,-5/2$ we see that $v^{\text{mp}}_{b\ell} \neq v^{\text{mp}}_M$. Given this, it is convenient to keep the factor \( 2k_B T/m \) as a typical velocity on dimensional grounds, but without assigning any specific physical meaning.  
\begin{table}[t!]
\renewcommand{\arraystretch}{2.8}
\caption{\label{T1.1}%
Fat-Tail Kappa Distributions from Eq (\ref{E14a}).}
\begin{ruledtabular}
\begin{tabular}{lc}
\noalign{\vskip -5pt}
 & \hspace{-10em}Distribution \\
\colrule
\noalign{\vskip -5pt}
 $\displaystyle \ell = 0$ & \hspace{-10em}FKK, \, $\kappa > 5/2$\\
 \noalign{\vskip -3pt}
  \multicolumn{2}{c}{$\displaystyle F(v) = n\left(\frac{m}{2\pi k_BT}\right)^{3/2} \frac{\Gamma\left(\kappa\right)}{\kappa^{3/2} \Gamma\left({\kappa} - \frac{3}{2}\right)}\left[1 +  \frac{m v^2}{2\kappa k_BT}\right]^{-\kappa}$}  \\[4pt]
\colrule
\noalign{\vskip -5pt}
$\displaystyle \ell=-1 $  &  \hspace{-10em}SKK, \, $\kappa > 3/2$ \\
\noalign{\vskip -3pt}
\multicolumn{2}{c}{$ \displaystyle F(v)=n\left(\frac{m}{2\pi k_BT}\right)^{3/2} \frac{ \Gamma\left(\kappa + 1\right)}{ \kappa^{3/2}\Gamma\left({\kappa} - \frac{1}{2}\right)}\left[1 + \frac{m v^2}{2\kappa k_BT}\right]^{-\kappa - 1}$} \\[4pt]
\colrule
\noalign{\vskip -5pt}
$\displaystyle \ell=-5/2$ &  \hspace{-10.5em}TKK, \, $\kappa > 0$  \\
\noalign{\vskip -3pt}
  \multicolumn{2}{c}{$\displaystyle F(v) = n\left(\frac{m}{2\pi k_BT}\right)^{3/2}  \frac{  \Gamma\left(\kappa+ \frac{5}{2}\right)}{ \kappa^{3/2}\Gamma\left({\kappa} + 1\right)} \left[1 +  \frac{m v^2}{2\kappa k_BT}\right]^{-\kappa - \frac{5}{2}}$} \\ [4pt]
  \colrule
\noalign{\vskip -5pt}
\end{tabular}
\end{ruledtabular}
\end{table}

\noindent\textit{4. On the effective temperature.} Let us now consider the MSKK as first discussed by Maxsimovic et al. \cite{Maksimovic1997}. Assuming the validity of SKK, their central idea was to preserve the  \,``Maxwellian form" of the average energy (per particle) that is, $\bar{\varepsilon}_{\kappa\ell}=3k_BT_{\kappa \,\ell}/2$.  To scrutinize this approach more closely, we reinterpret their proposal from a broader perspective, based on the general ($\kappa,\,\ell$) VDF. From Table II, the average energy density $\bar{\varepsilon}_{\kappa \ell}$  can be written as:
\begin{equation}\label{Eq20}
 \bar{\varepsilon}_{\kappa\ell}= \frac{3k_B T}{2} \frac{\kappa}{\kappa - \ell - 5/2 }\equiv{\frac{3}{2}k_BT_{\kappa\ell}}, 
\end{equation}
where $T_{\kappa\ell}$ is an effective temperature defined by:
\begin{equation} \label{Eq21}
  T_{\kappa\ell} = T \frac{\kappa}{\kappa - \ell-5/2 }.
\end{equation} 
Now, by inserting (\ref{Eq21}) into (\ref{Eqn24}) we obtain: 
\begin{eqnarray}\label{Eqn22}
     F(v) &=& n\left(\frac{m}{2\pi k_B T_{\kappa\ell}(\kappa - \ell- 5/2  ) }\right)^{3/2} \frac{ \Gamma\left(\kappa - \ell\right)}{ \Gamma\left(\kappa - \ell - \frac{3}{2}\right)}\nonumber \\ &\times& \left[1 + \frac{ m v^2}{2k_B T_{\kappa \ell}(\kappa - \ell- 5/2)}\right]^{-(\kappa - \ell)},
\end{eqnarray}
which represents a general modified VDF for arbitrary values of the pair ($\kappa, \ell$). Hence, the equivalent typical velocity may be defined as: 
\begin{equation}\label{E23}
   v_{\kappa\ell}^2 = \frac{2k_B T_{\kappa \ell}}{m} \,\frac{\kappa-\ell -5/2}{\kappa}. 
\end{equation} 
This means that any VDF using $T_{\kappa\ell}$ as given by (\ref{Eq21}) can be called a modified version of the original ($\kappa, \ell$) VDF.  For instance, for $\ell=-1$ we have the MSKK distribution:
\begin{eqnarray}\label{Eqn26}
     F(v) &=& n\left(\frac{m}{2\pi k_B T_{\kappa}(\kappa -  3/2  ) }\right)^{3/2} \frac{ \Gamma\left(\kappa +1\right)}{ \Gamma\left(\kappa - \frac{1}{2}\right)}\nonumber \\ &\times& \left[1 + \frac{ m v^2}{2k_B T_{\kappa}(\kappa - 3/2)}\right]^{-(\kappa - 1)},
\end{eqnarray}
where $T_{\kappa} = T[\kappa/(\kappa-3/2)]$. When written in terms of $v_{\kappa}^2=2k_B T_{\kappa}[(\kappa-3/2)/m\kappa]$, this is exactly the distributions defined by Maksimovic et al.  (Table \ref{T0}).

However, for FKK distributions ($\ell = 0$), the effective temperature in Eq. (\ref{Eq21}) is $T_{\kappa} = T[\kappa/(\kappa-5/2)]$, thus generating a new VDF (named here MFKK). A surprising aspect of the present investigation is to recognize that an ``effective  $T_{\kappa\,\ell}$ temperature" can always be introduced by choosing the pair ($\kappa,\,\ell$), which is determined by the mean energy values. 

Nevertheless, for $\ell=-5/2$, the mean energy per particle and even the thermostatic pressure remain completely independent of the free parameters (see Table II and IV). This interesting property allows the construction of TKK distributions with any suitable $\kappa$-index, all of them sharing the same average energy per particle and pressure, as given in the Maxwellian case. As we shall see below, this temperature $T$ is in agreement with the suprathermal nature of the data generating the fat-tail distribution. This means that, potentially, $k_{B}T$ is the most relevant scale to measure energy. In other words, it is not necessary to know how $T$, $\kappa$ and $\ell$ for determining the energy density, since only $T$ is required for TKK. Hence, it seems that only in this special case, the somewhat artificial ($\kappa,\ell$)-dependent temperature does not need to be introduced. Naturally, one may also argue that if the measured average energy $\bar{\varepsilon}$ should be distributed among the parameters $T$, $\kappa$ and $\ell$ in agreement with Eqs. (\ref{Eq20}) and (\ref{Eq21}), the physical temperature $T$ should be smaller than the Maxwellian result since $\kappa/(\kappa -\ell -5/2)\geq 1$.

\begin{figure}[t!]
   \centering
   \includegraphics[width=0.5\textwidth]{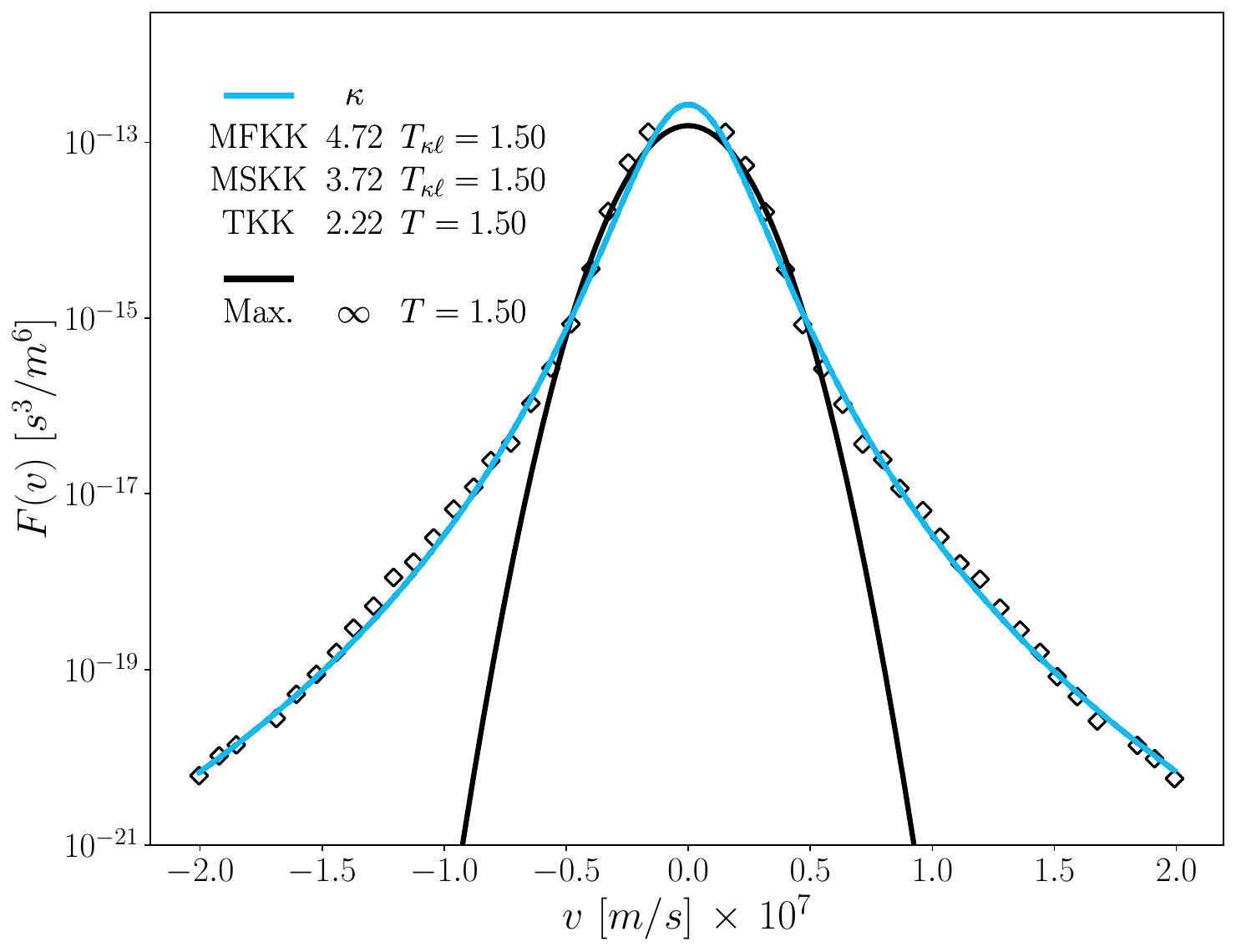}
   \vspace{-10pt}
      \caption{Electron VDFs in the slow solar wind. The blue line represents the adjustment of the modified Kappa models for $\ell = 0$ \,\text{(MFKK)},  $\ell=-1$ \,\text{(MSKK)}, and also the \text{(TKK)} model with $\ell=-5/2$. The solid black line illustrates the Maxwellian distribution. Note that all distributions predicts the same temperature in units of $10^{5} \text{K}$.  However, only the TKK distribution ($\ell= -5/2$), fits the data with the physical temperature $T$ (see text for more details).} 
    \label{f1}
\end{figure}

In Figure \ref{f1} we confront the modified VDFs with TKK and Maxwell distributions based on the data used by Salem et al.  \cite{Salem2023}. The $\bar{\varepsilon}_{\kappa \, \ell}$ is fixed by the total average energy per particle measure from the data points. When evaluated on a logarithmic scale, all fits coincide with the solid blue line, given the concentration
$n = 7$ cm$^{-3}$. In the Maxwellian case, solid black line,
$n = 8$ cm$^{-3}$. This symmetry occurs because by fixing
$\ell$, the parameter $\kappa$ is adjusted so that the combination $\kappa-\ell$ remains constant in Eqs. (\ref{Eqn24}) and (\ref{Eqn22}). As a result,
for any modified functions, the factor $T_{\kappa \,\ell}(\kappa - \ell - 5/2)$ also remains constant with the same temperature $T_{\kappa \,\ell} = 1.5 \times 10^5 \,\text{K}$. Moreover, the TKK distribution provides a remarkable fit with $T = 1.5 \times 10^5\, \text{K}$, identical to the Maxwellian temperature. In both cases, $T$ is independent of $\kappa$. Together, all these results indicate that $T$ represents the physical temperature of the system, suggesting that the effective temperature ($T_{\kappa \,\ell}$) may no longer be necessary.


Furthermore, $\ell$ also imposes constraints to prevent physical quantities from becoming negative or complex. It is easy to verify that the most restrictive constraint is given by the average energy per particle in Table \ref{T1}, which can be written for $\kappa$:
\begin{equation}\label{E21} 
\begin{array}{ll}
\kappa > \ell + 5/2, 
\end{array} 
\end{equation} 
for $\ell = 0$ (FKK) and $\ell =-1$ (SKK, MSKK) we have $\kappa >  5/2$ and $\kappa > 3/2$, respectively.
Interestingly, for $\ell = -5/2 $ there are no constraints on the physical quantities (at least up to the second moment of the distribution function). Hence, since TKK fits the data for any $k>0$ and $T$ is the physical temperature, it seems natural to argue that it would be the most physically appealing distribution of the ($\kappa,\ell$) class derived here. 

\noindent{\it 5. Final Comments.} The extended Neo-Maxwellian approach developed in this article provides a unified framework for modeling ideal gas systems with non-Gaussian velocity distributions, now characterized by two free parameters ($\kappa,\ell$). Such parameters describe two large classes of \textit{short} and \textit{fat-tails}, $\kappa$-distributions proposed in the literature, with normalization and thermodynamic quantities adjusted accordingly. Even unknown Kappa distributions can be described for finite but arbitrary values of $\ell$. This approach is in perfect agreement with homogeneity, isotropy, and thermal equilibrium. It clarifies long-standing kinetic subtleties, and also some ``mysteries" or ambiguities related to the $\ell$-numbers in the context of Kappa-type distributions. The special case $\ell = -5/2$ appears to be the most appropriate choice, for which $T$ becomes the only relevant kinetic energy scale, as in the Maxwellian case, allowing the measurement of suprathermal energy without introducing any effective temperature $T_{\kappa}$. Note also that for $\kappa^{-1} = q-1$ and $\ell\neq 0$ this approach provides a simple realistic extension for Tsallis power law distributions \cite{Silva1998,Tsallis1988,Lima2001,Lima2000b}  

 It is worth mentioning that Maxwell factorization condition is similar to the Boltzmann hypothesis of `molecular chaos'. This hypothesis played a fundamental role in the standard  Boltzmann’s kinetic approach to the equilibrium Maxwell’s velocity distribution when the source of entropy is nullified. As discussed a couple of years ago for short-tailed distributions \cite{Lima2020}, the present  ($\kappa,\ell$) framework can also be extended to the nonadditive entropic regime, by adopting a suitably formulation for `molecular chaos' hypothesis tailored for finite systems.
 
Another aspect that still warrants deeper investigation is the formulation of a proper deformed entropy, $S_{\kappa\,\ell}$, and its connection to thermodynamics.  In principle, addressing this problem may shed light on the nature of the correlations associated with  ($\kappa,\,\ell$) fat-tailed velocity distributions, and help to determine the appropriate value of the $\ell$-parameter. This topic will be discussed in a forthcoming communication.
 

\vspace{5pt}



\noindent\textbf{Acknowledgments:} JASL is partially supported by CNPq (310038/2019-7) and FAPESP under grants (LLAMA project, 11/51676-9 and 24/02295-2). MHB is supported by FAPESP/CNPq (24/14163-3).
\newpage
\appendix
\section{Short-Tail Kappa Distributions} In this Appendix  the results for short-tailed distributions will be discussed. To begin with, let us obtain the physical quantities as a function of the pair ($\kappa,\,\ell$). From equation (\ref{E14b}) we set $ b = |\kappa|^{-1}$, thereby obtaining:
\begin{equation}\label{Eqn25}
         F(v) = n\left(\frac{\beta m}{2\pi}\right)^{\frac{3}{2}} \frac{ \Gamma\left(|\kappa| + \ell + \frac{5}{2}\right)}{ |\kappa|^{\frac{3}{2}}\Gamma\left(|\kappa| + \ell + 1\right)} \left[1 - \frac{ \beta m v^2}{2|\kappa| }\right]^{|\kappa| + \ell},
\end{equation}
where $\beta = (k_B T)^{-1}$.

Now, it is easily seen that the short-tailed family, which is the counterpart of the fat-tailed distributions in Table \ref{T1.1}, is obtained by taking the same values $\ell = 0 $ (FKK), $ \ell = -1 $ (SKK, MSKK), $\ell=-5/2$ (TKK). 

In Table \ref{T1S} we present the short-tailed results from the typical averaged quantities in terms of the pair of parameters ($\kappa,\ell$). These results should be compared with the fat-tailed case presented in Table II. 

In Table \ref{T2S} the final forms of short-tailed distributions are given.  These distributions are generalizations for negative values of the $\kappa$-index (for more details, see ref. \cite{Leubner2004}). Note also that for $\ell = 0$ (FKK) and $\ell=-1$ (SKK, MSKK) we have $|\kappa| > 0 \,(\kappa < 0)$. For $\ell=-5/2$ (TKK) the constraint will be $|\kappa| > 3/2  \, \,(\kappa <-3/2)$ .
\\

\begin{table}[h!]
\renewcommand{\arraystretch}{2.8}
\caption{\label{T1S} Quantities for Short-tailed distributions ($\kappa<0$)}
\begin{ruledtabular}
\begin{tabular}{cc}
$\bar{\varepsilon}_{\kappa\ell} =  \displaystyle\bar{\varepsilon}_M \left( \frac{|\kappa|}{|\kappa|+\ell +5/2   } \right)$ 
& $P_{b\ell} =  \displaystyle P_M \left(\frac{|\kappa|}{|\kappa|+\ell +5/2   }\right)$ \\[6pt]
$v^{\text{rms}} =  \displaystyle v^{\text{rms}}_M\sqrt{\frac{|\kappa|}{|\kappa|+\ell +5/2   }} $ 
& $\bar{v} =  \displaystyle\bar{v}_M \frac{|\kappa|^{1/2}\Gamma\left(|\kappa|+\ell+ 5/2\right)}{ \Gamma\left(|\kappa|+\ell +3\right)} $ \\[6pt]
$v^{\text{mp}} =  \displaystyle v^{\text{mp}}_M\sqrt{\frac{|\kappa|}{|\kappa|+\ell +1}} $ \\[3pt]
\end{tabular}
\end{ruledtabular}
\end{table}
\begin{table}[h!]
\renewcommand{\arraystretch}{2.8}
\caption{\label{T2S}
Short-Tailed Kappa Distributions from Eq. (\ref{E14b}). FKK (i), SKK (ii), TKK (iii) and  MSKK (iv).}
\begin{ruledtabular}
\begin{tabular}{l}
$ \overset{(i)}{F}(v) = \displaystyle n\left(\frac{m}{2\pi k_BT}\right)^{3/2} \frac{ \Gamma\left(|\kappa| + \frac{5}{2}\right)}{ |\kappa|^{3/2}\Gamma\left({|\kappa|+1} \right)}\left[1 -  \frac{m v^2}{2|\kappa| k_BT}\right]^{|\kappa|} $ \\ [6pt]
 $  \overset{(ii)}{F}(v) = \displaystyle n\left(\frac{m}{2\pi k_BT}\right)^{3/2} \frac{ \Gamma\left(|\kappa| + \frac{3}{2}\right)}{ |\kappa|^{3/2}\Gamma\left({|\kappa|} \right)}\left[1 -  \frac{m v^2}{2|\kappa| k_BT}\right]^{|\kappa|-1} $ \\ [6pt]
 $  \overset{(iii)}{F}(v) =\displaystyle n\left(\frac{m}{2\pi  k_BT}\right)^{3/2}  \frac{ \Gamma\left(|\kappa|\right)}{ |\kappa|^{3/2}\Gamma\left({|\kappa|-\frac{3}{2}}\right)} \left[1 -  \frac{m v^2}{2|\kappa| k_BT}\right]^{|\kappa| - \frac{5}{2}} $ \\ [6pt]
 $  \overset{(iv)}{F}(v) = \displaystyle n\left(\frac{m}{2\pi v^2_{\kappa}}\right)^{3/2}  \frac{ \Gamma\left(|\kappa|+ \frac{3}{2}\right)}{|\kappa|^{3/2}\Gamma\left({|\kappa|}\right)} \left[1 -  \frac{m v^2}{2|\kappa| v^2_{\kappa}}\right]^{|\kappa| - 1} $ \\
\end{tabular}
\end{ruledtabular}
\end{table}


\end{document}